\documentclass[journal=jacsat,manuscript=article]{achemso}

\usepackage{graphicx}
\usepackage{dcolumn}
\usepackage{bm}
\usepackage{hyperref}
\usepackage{color,soul} 

\author{Xuelei Sui}

\affiliation
{Department of Physics and State Key Laboratory of Low-Dimensional Quantum Physics, Tsinghua University, Beijing 100084, China}

\author{Jianfeng Wang}
\email{wangjf@csrc.ac.cn}
\affiliation{Beijing Computational Science Research Center, Beijing 100193, China}

\author{Wenhui Duan}
\email{dwh@phys.tsinghua.edu.cn}
\affiliation
{Department of Physics and State Key Laboratory of Low-Dimensional Quantum Physics, Tsinghua University, Beijing 100084, China}
\alsoaffiliation
{Institute for Advanced Study, Tsinghua University, Beijing 100084, China}

\title{Prediction of Stoner-Type Magnetism in Low-Dimensional Electrides}

\abbreviations{FM, AFM, $E_F$}

\begin{document}


\begin{tocentry}

\begin{center}

\includegraphics[width=7.5cm]{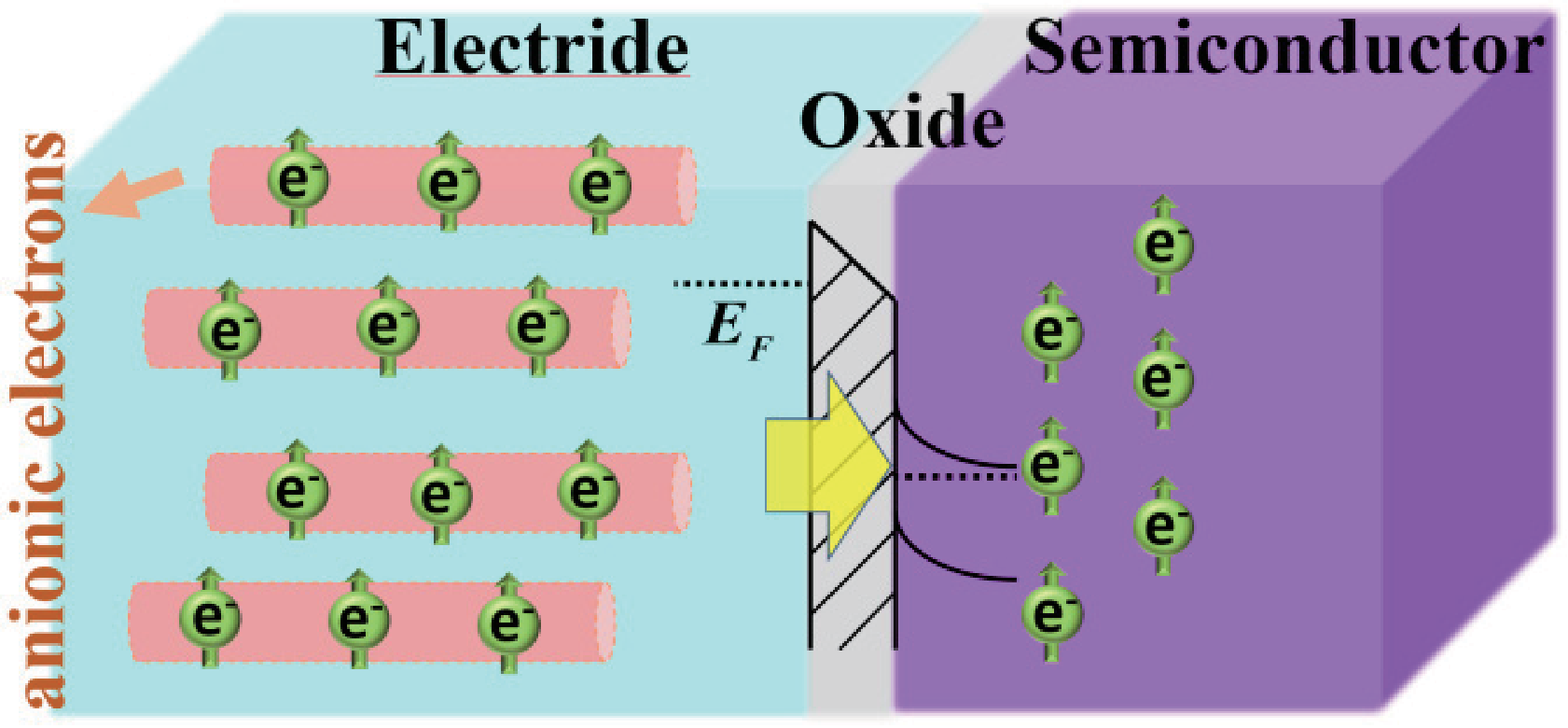}

\end{center}

\end{tocentry}

\begin{abstract}
Electrides are special ionic solids with excess cavity-trapped electrons serving as anions. Despite the extensive studies on electrides, the interplay between electrides and magnetism is not well understood due to the lack of stable magnetic electrides, particularly the lack of inorganic magnetic electrides. Here, based on the mechanism of Stoner-type magnetic instability, we propose that in certain electrides the low-dimensionality can facilitate the formation of magnetic ground state because of the enhanced density of states near the Fermi level. To be specific, $A_5B_3$ ($A$ = Ca, Sr, Ba; $B$ = As, Sb, Bi) (1D), Sr$_{11}$Mg$_2$Si$_{10}$ (0D), Ba$_7$Al$_{10}$ (0D) and Ba$_4$Al$_5$ (0D) have been identified as stable magnetic electrides with spin-polarization energies of tens to hundreds of meV per formula unit. Especially for Ba$_5$As$_3$, the spin-polarization energy can reach up to 220 meV. Furthermore, we demonstrate that the magnetic moment and spin density mainly derive from the interstitial anionic electrons near the Fermi level. Our work paves a way to the searching of stable magnetic electrides and further exploration of the magnetic properties and related applications in electrides.
\end{abstract}

\section{Introduction}
Electrides are special ionic solids, in which there are excess valence electrons from the viewpoint of formal valence.\cite{first_organic,electrides,anionic,anion} The excess electrons are trapped at the cavities which are formed by atomic voids, and thus serve as anions.\cite{highpressure} Due to the loose bonding to the cores, these anionic electrons show low work function\cite{work_function_Ca24Al28O64} and high mobility,\cite{mobility} making the electrides promising catalysts,\cite{catalyst1,catalyst2} electron emitters,\cite{emmiter} and battery anodes,\cite{injection,anode} etc.\cite{JPCL_Optical,JPCL_Battery} The earliest researches of solid electrides focused on the organic materials, which, however, are unstable at room temperature and sensitive to the atmosphere.\cite{first_organic} In 2003, the first stable inorganic electride, [Ca$_{24}$Al$_{28}$O$_{64}$]$^{4+}${$\cdot$}4e$^-$, was prepared in experiment.\cite{Ca24Al28O64,calculateCa24Al28O64} In recent years, Ca$_2$N has been demonstrated to be an electride by electron transport measurement and since then inorganic electrides have sprung up quickly.\cite{Ca2N,Ca2N2,Ca2N3,2Delectride,JPCL_1D} According to the dimensionality of the anionic electrons confinement, e.g., zero-dimensional (0D) cavities, one-dimensional (1D) channels and two-dimensional (2D) interlayers, the electrides can be conveniently classified. A large amount of electrides with different confinement dimensionalities have been identified recently by high-throughput calculations from all known inorganic materials.\cite{Throughput} Therefore, electride has become an active field of solid state chemistry and material science.

The magnetic electrides stand out for the significant potential application in spintronics. For instance, the spin-polarized anionic electrons with low work function can be used for spin injection.\cite{spininjection} However, the explorations on the properties and applications related to the magnetism, which mainly comes from the trapped anionic electrons, are limited due to the lack of stable magnetic electrides. Considering the itinerant nature of the electride electrons, the magnetic instability is believed to be originated from the Stoner's mechanism, i.e., the ferromagnetism instability caused by the high peak of electronic density of states (DOS) near the Fermi level ($E_F$).\cite{Stoner} Early attempts have been made in the alkali metal organic antiferromagnetic electrides, where the magnetism cannot be well utilized and manipulated.\cite{AFM1,AFM2} For the inorganic electrides, a 2D magnetic electride Y$_2$C has been studied very recently, which can be easily exfoliated into thin films and benefit the 2D spintronic devices.\cite{Y2C1} However, the calculated spin-polarization energy of the ferromagnetic state is low, and no static magnetic order of the anionic electrons has been observed in experiments even at low temperature.\cite{Y2C_calcu,Y2C2} To stabilize its magnetic state, the sample quality needs improving and assistant external methods are required.\cite{electric} Therefore, stable inorganic magnetic electrides, especially those with ferromagnetic ground state and large spin-polarization energy, are urgently needed for the future magnetic applications.

In this work, based on the mechanism of Stoner-type magnetic instability, we propose that the magnetic ground state could be realized in 1D and 0D electrides due to the possibly high DOS near $E_F$. By searching all the reported inorganic electride materials with this guiding principle, we identify four classes of stable magnetic electrides: $A_5B_3$ ($A$ = Ca, Sr, Ba; $B$ = As, Sb, Bi) (1D), Sr$_{11}$Mg$_2$Si$_{10}$ (0D), Ba$_7$Al$_{10}$ (0D) and Ba$_4$Al$_5$ (0D). The anionic electrons are located in the interstitial sites of 1D channels or 0D atomic voids shown by the charge densities, indicating the character of electrides. Importantly, these electrides all have magnetic ground states with either ferromagnetic (FM) or antiferromagnetic (AFM) orders, and the spin-polarization energies can be up to hundreds of meV. The peak of DOS around $E_F$ in the nonmagnetic calculations splits and shifts when performing the spin-polarized calculations, confirming the Stoner-type instability. Spatial spin density distribution shows that the magnetic moment is primarily originated from the interstitial anionic electrons.

\section{Methods}

First-principles calculations within the framework of density functional theory (DFT) are performed by the Vienna $ab$ $initio$ simulation package (VASP).\cite{VASP} The Perdew-Burke-Ernzerhof (PBE) generalized gradient approximation (GGA) of the exchange-correlation functional is used.\cite{PBE} The ionic potentials of all atoms are described by projector augmented wave (PAW) method\cite{PAW}. The plane wave basis with a kinetic energy cutoff of 400 eV and fine $k$-point mesh denser than 0.1 \AA$^{-1}$ are adopted, yielding well-converged total energies. All the structures are fully relaxed with van der Waals correction (DFT-D2 method) until the residual force acting on each atom is lower than 0.001 eV/\AA.\cite{VDW} We find that the relaxed lattice constants of all materials are consistent with the experimental values. Spin-polarized calculations for both magnetic and nonmagnetic states are employed to get the magnetic ground states and electronic properties. The initial magnetic bias is applied on Ba atoms neighboring to the voids which are allowed to optimize. For nonmagnetic state, the initial magnetic bias is set to be zero, and the energies calculated using such method are almost the same as the results of non-spin-polarized calculations. An overestimation of the nonmagnetic energy may be caused if considering the static correlation effect for the spin-degenerate system.\cite{error DFT} Heyd-Scuseria-Ernzerhof (HSE06) hybrid functional calculations with the same cut-off energy and $k$-mesh as the PBE calculation are carried out to check the results.\cite{HSE} Hole doping is performed by reducing the number of electrons and meanwhile assuming a homogeneous background charge, which is a universal method for the doped system.\cite{dope}

\section{Results and discussion}

\subsection{Stoner-type magnetism in electride}

To search for the stable magnetic electrides, we start from the Stoner's mechanism, with which the origin of itinerant electron magnetism can be successfully explained.\cite{Stoner} For a metallic system with itinerant electrons, if it has DOS peak at $E_F$ or large exchange interaction, the FM state will be likely formed in order to lower the total energy. In other words, the FM instability occurs when $D$($E_F$)$I$ $>$ 1, where $D$($E_F$) and $I$ are DOS at $E_F$ and the effective exchange interaction parameter, respectively. For magnetic electrides we concern here, the magnetism is mainly contributed by the interstitial electrons. As to the exchange interaction $I$ which is complicated to handle, the values for different electrides might be similar because of the $s$-like feature of the anionic electron.\cite{Y2C_calcu} Therefore, the $D$($E_F$), which varies greatly for electrides of different dimensionalities, plays a decisive role in the FM instability. The itinerant anionic electrons are trapped in the cavities, leading to the possibly large DOS near $E_F$ due to the quantum confinement effect.

\begin{figure} [tbp]
\centering
 \includegraphics[width=0.8\textwidth]{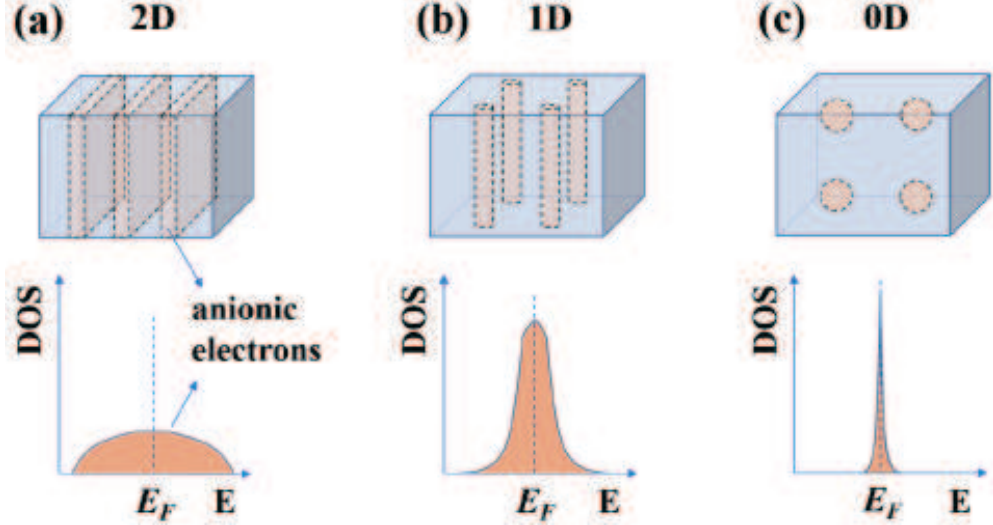}\\
 \caption{(Color online) Schematic diagrams of the structure and DOS for (a) 2D, (b) 1D and (c) 0D electrides. The orange regions represent the trapped anionic electrons, and the dashed lines in the DOS label the Fermi level.}
 \label{fig:Figure1}
\end{figure}

As shown in Figure 1a, the 2D electride has layered crystal structure, and the anionic electrons are confined in the interlayer space. The broad intralayer band dispersion results in a relatively small DOS near $E_F$, which may be one of the reasons for the weak magnetism and low energy reduction in 2D electride Y$_2$C.\cite{Y2C_calcu} For the case of 1D or 0D electrides, the anionic electrons reside in the fibrous channels or point cavities with 1D or 0D energy dispersion [Figure 1b,c]. A large peak of DOS is likely to generate near $E_F$ due to the quantum confinement effect, satisfying the condition of the Stoner-type instability. We underline that the AFM state may also be formed due to the complicated band structure and exchange interaction in electrides. Therefore, in the 1D and 0D electrides, the magnetic ground states can form thank to the possibly large DOS near $E_F$. For instance, the AFM ground states of the early studied organic electrides can be attributed to the 0D confinement. It should be emphasized that the model only gives the ideal situation, i.e., the anionic electrons are fully trapped in cavities without interacting with the orbital electrons. In real materials, the orbital electrons are likely to hybridize with the anionic electrons, affecting the localization of the anionic electrons and the spin splitting. That is to say, the $D$($E_F$) of electrides depends on both the dimensionality of the confinement and the degree of hybridization with orbital electrons. Here, we mainly focus on the former factor and leave the latter for future discussion.

Recently, a high-throughput identification of electrides from all known inorganic materials was reported,\cite{Throughput} and thus we limit our search of magnetic electrides to the demonstrated 69 inorganic electride candidates. There are 51 1D and 0D electride materials, among which 12 materials are first found to be magnetic, i.e. II-V compounds $A_5B_3$ ($A$ = Ca, Sr, Ba; $B$ = As, Sb, Bi) (1D), Sr$_{11}$Mg$_2$Si$_{10}$ (0D), Ba$_7$Al$_{10}$ (0D) and Ba$_4$Al$_5$ (0D). It is worth noting that some electrides also have magnetic moment, but their magnetization is contributed by the orbital electrons. Since it is the anionic electrons that mainly contribute to the chemical activity, here we only recognize the materials with spin density located on the anionic electrons as magnetic electrides. In the following, we will present detailed calculation results and analyses for the identified inorganic magnetic electrides.

\subsection{1D magnetic electrides}

\begin{figure} [tbp]
\centering
 \includegraphics[width=0.6\textwidth]{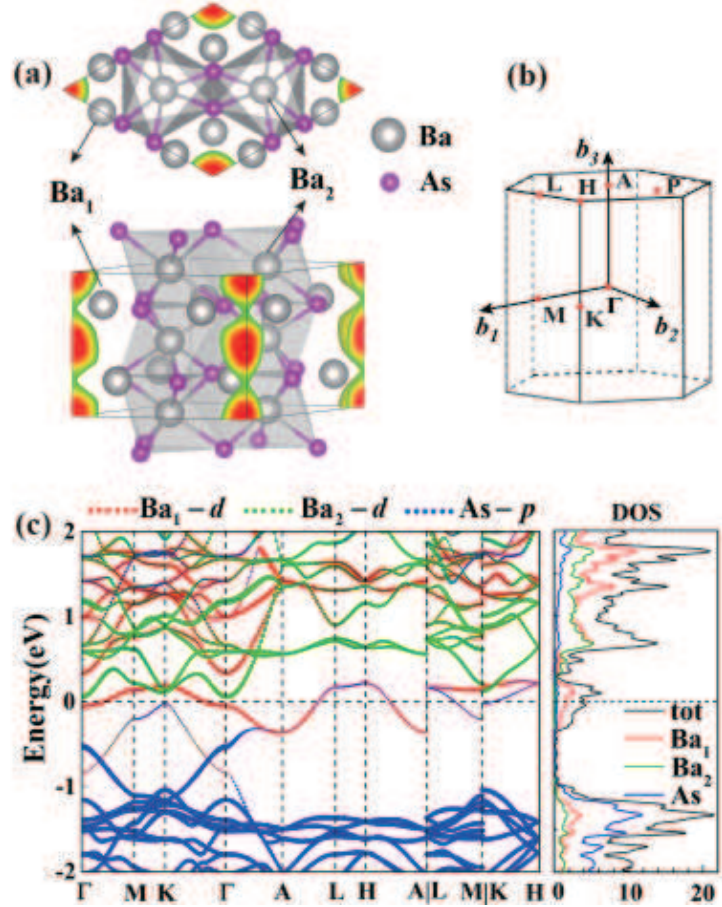}\\
 \caption{(Color online) (a) Crystal structure of Ba$_5$As$_3$ with upper (lower) panel for the top (side) view. The gray and pink spheres indicate Ba and As atoms respectively. Two inequivalent Ba sites are labeled by Ba$_1$ and Ba$_2$. The partial charge density around $E_F$ ($-$0.3 eV to 0.3 eV) denotes the anionic electrons (red regions). Isosurface is set at 0.02 $e$/\AA$^3$. (b) Brillion zone of Ba$_5$As$_3$. The high symmetry points are labeled by red dots. (c) Band structure with atomic orbital projections and partial DOS from NM calculations. The dashed horizontal line labels the Fermi level.}
 \label{fig:Figure2}
\end{figure}

II-V compounds $A_5B_3$ ($A$ = Ca, Sr, Ba; $B$ = As, Sb, Bi) have hexagonal structures with space group of $P6_3/mcm$ (Figure 2a). We take Ba$_5$As$_3$ as an example to illustrate the structure and electronic properties. The crystal structure is shown in Figure 2a. The calculated lattice constants are $a$ = $b$ = 9.56 {\AA} and $c$ = 7.81 \AA, consistent with the experimental values ($a$ = $b$ = 9.49 {\AA}, $c$ = 7.90 \AA). There are two inequivalent Ba sites, Ba$_1$ and Ba$_2$, in the unit cell. Three out of five Ba atoms (Ba$_2$) sit at the center of the face-sharing octahedra, while the remaining two Ba atoms (Ba$_1$) form a 1D cavity channel together. In the regard of the formal valence states of Ba$^{2+}$ and As$^{3-}$, there is one excess electron for Ba$_5$As$_3$. Obviously, Ba$_5$As$_3$ meets two common conditions for electrides: excess electrons and interstitial cavities. To see whether these excess electrons are trapped in the cavities, we first demonstrate the results of non-spin-polarized calculations. The band structure along high-symmetry lines in Brillouin zone [Figure 2b] is shown in Figure 2c, indicating the metallic feature of Ba$_5$As$_3$. For the energy bands near $E_F$, which consist of the excess electrons, there is almost no contribution from the atomic orbitals except for a small projection of Ba$_1$ atoms. The partial charge densities of these bands in the energy interval from $-$0.3 eV below to 0.3 eV above $E_F$ further show that the excess electrons are confined in the 1D cavity channels [Figure 2a]. Therefore, Ba$_5$As$_3$ is an electride with excess electrons serving as anions. The unit cell containing two molecular formulas can be written as [Ba$_{10}$As$_6$]$^{2+}$ $\cdot$ 2e$^{-}$. Moreover, the nature of the anionic electrons can also be revealed by the comparison of the total DOS and partial DOS [right panel of Figure 2c]. The small atomic orbital occupations around $E_F$ indicate that the states around $E_F$ are mainly located in the interstitial sites.

As a result of the non-spin-polarized calculations, Ba$_5$As$_3$ is a 1D electride with the anionic electrons trapped in 1D channels. Due to the small energy dispersion in $xy$ plane, there is a peak of DOS near $E_F$ composed of the anionic electrons [Figure 2c], which may give rise to the Stoner-type instability. Next, we show the results of spin-polarized calculations.

\begin{figure} [tbp]
\centering
 \includegraphics[width=0.75\textwidth]{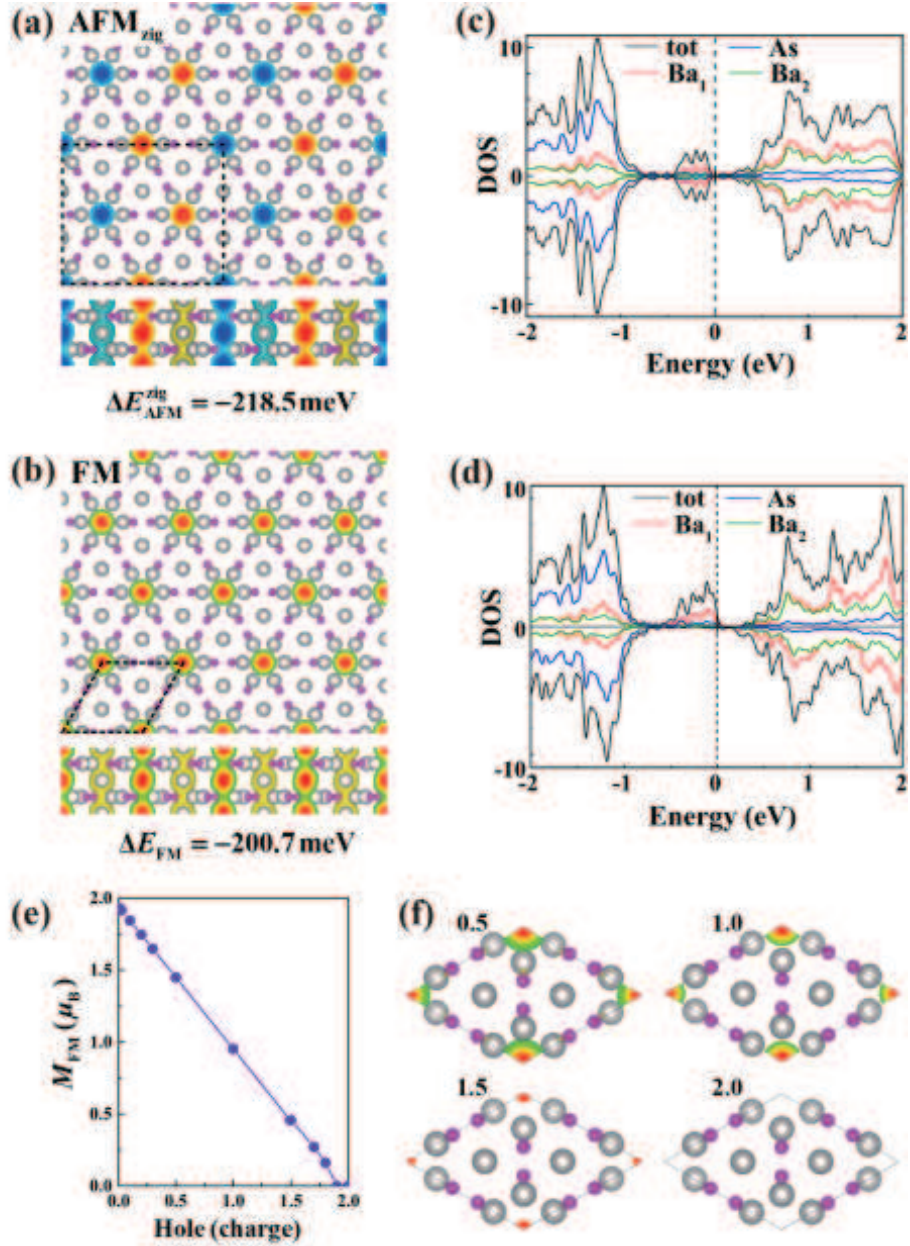}\\
 \caption{(Color online) (a,b) Spin density distributions with isosurface at 0.02 $\mu_B$/\AA$^3$ for zigzag AFM (AFM$_{\rm zig}$) and FM Ba$_5$As$_3$. Upper (lower) panel is the top (side) view. Red and blue regions represent the majority and minority spin densities, respectively. $\Delta E_{\rm AFM}^{\rm zig}$ and $\Delta E_{\rm FM}$ below are the energies of AFM$_{\rm zig}$ and FM states relative to the NM state. The black dashed line labels the unit cell of the magnetic states. (c,d) Total and partial DOS for AFM$_{\rm zig}$ and FM states. (e,f) The magnetic moment and spin density (isosurface at 0.01 $\mu_B$/\AA$^3$ ) for FM state as functions of the number of the removed electrons.}
 \label{fig:Figure3}
\end{figure}

Here, three magnetic configurations are taken into consideration: FM, zigzag AFM (AFM$_{\rm zig}$) and collinear AFM (AFM$_{\rm coll}$) states. Though the anionic electrons don't belong to any atoms, the initial magnetic bias can be applied to the nearest Ba$_1$ atoms neighboring to the voids, which is a successful approach used in the calculation of organic magnetic electrides.\cite{calculatemagnetic,PCCP2018} The calculated total energies show that the AFM$_{\rm zig}$ state is the most stable magnetic order [Figure 3a], with the energy lower than the NM state by 218.5 meV per unit cell. While the energy of the FM state is 200.7 meV lower than the NM state [Figure 3b], slightly higher than the AFM$_{\rm zig}$ state. The energy difference between AFM$_{\rm coll}$ and NM is $-$217.4 meV (data are shown in Table 1). It should be noted that the spin-polarization energy is much larger than that of 2D electride Y$_2$C, which is only tens of meV.\cite{Y2C_calcu} This result indicates the strong stability of magnetism in Ba$_5$As$_3$. Figure 3a,b show the spatial spin density distributions, which mainly concentrate in the interstitial 1D channels for both AFM$_{\rm zig}$ and FM states. So we can conclude that the magnetism is primarily induced by the anionic electrons located in the 1D chain. Meanwhile, the magnetic moment of the FM state is around 2 $\mu_B$, in good accordance to the two excess electrons per unit cell. In addition, the peak of DOS splits and shifts away from $E_F$ [Figure 3c,d] in the magnetic state. (The band structures for these magnetic states can be referred to Figure S1.) Whether the magnetic ground state is FM or AFM depends on the exchange interaction of the anionic electrons, which is complex and beyond this work. Ruderman-Kittel-Kasuya-Yosida (RKKY) interaction may also be one of the magnetic coupling mechanisms for the anionic electrons, i.e., the anionic electrons interact with each other through conducting $d$ electrons of Ba.

To further confirm that the anionic electrons play the main role in the magnetism, we artificially reduce the number of excess electrons of Ba$_5$As$_3$, which can be achieved by hole-type doping or gating in 2D structure. Figure 3e,f show the magnetic moment and spin density for the FM state as functions of the number of the removed electrons. As the excess electrons are taken away gradually, the magnetic moment decreases linearly, and finally to zero when two anionic electrons in a unit cell are completely removed. The spin density map [Figure 3f] also shows that the magnetization density in the 1D cavity is reduced accordingly and finally disappears. Therefore, it is reasonable to conclude that the anionic electrons in the 1D channel dominate the magnetism of Ba$_5$As$_3$.

\renewcommand\arraystretch{1.2}\setlength{\tabcolsep}{6pt}
\begin{table}[tbp]
	\caption{Calculated magnetic properties and spin-polarization energies of electrides $A_5B_3$ ($A$ = Ca, Sr, Ba; $B$ = As, Sb, Bi), where the magnetic moments $M_{\rm FM}$ are for the FM states (in unit of $\mu_B$) and $\Delta E_{\rm FM}$, $\Delta E_{\rm AFM}^{\rm coll}$, $\Delta E_{\rm AFM}^{\rm zig}$ (in units of meV) are the energies of FM, AFM$_{\rm coll}$ and AFM$_{\rm zig}$ states relative to the NM state. The energies of AFM$_{\rm coll}$ and AFM$_{\rm zig}$ states are similar for each electride, so we only give the relative energy difference between FM and AFM$_{\rm zig}$ states, denoted by $\Delta E_{\rm rel}$. There are two molecular formulas in one unit cell. For each compound, the ground magnetic state is labeled by *.}
	\label{tb:tb1}
	\begin{tabular}{cccccc}
		\hline
		\hline
		System  &$M_{\rm FM}$  &$\Delta E_{\rm FM}$   &$\Delta E^{\rm coll}_{\rm AFM}$  &$\Delta E^{\rm zig}_{\rm AFM}$ &$\Delta E_{\rm rel}$ \\
		\hline
		Ca$_{10}$As$_6$  &$1.63$   &$-82.0$\textsuperscript{*}    &$-54.4$   &$-54.5$  &$-27.5$ \\
		Ca$_{10}$Sb$_6$  &$1.55$   &$-57.7$\textsuperscript{*}    &$-1.1$    &$-1.2$  &$-56.5$ \\
        Ca$_{10}$Bi$_6$  &$1.46$   &$-46.0$\textsuperscript{*}    &$29.1$    &$28.6$  &$-74.6$ \\
        Sr$_{10}$As$_6$  &$1.78$   &$-101.0$   &$-102.1$\textsuperscript{*}  &$-101.4$ &$0.4$ \\
        Sr$_{10}$Sb$_6$  &$1.83$   &$-110.1$\textsuperscript{*}   &$-86.0$   &$-86.3$ &$-23.8$ \\
        Sr$_{10}$Bi$_6$  &$1.76$   &$-111.8$\textsuperscript{*}   &$-74.0$   &$-73.8$ &$-38.0$ \\
        Ba$_{10}$As$_6$  &$1.93$   &$-200.1$   &$-217.4$  &$-218.5$\textsuperscript{*} &$18.4$ \\
        Ba$_{10}$Sb$_6$  &$1.95$   &$-179.2$   &$-186.1$  &$-186.4$\textsuperscript{*} &$7.2$ \\
        Ba$_{10}$Bi$_6$  &$1.91$   &$-163.2$\textsuperscript{*}   &$-161.5$  &$-161.5$ &$-1.7$ \\
		\hline
	\end{tabular}\\
\end{table}

Other II-V compounds $A_5B_3$ ($A$ = Ca, Sr, Ba; $B$ = As, Sb, Bi) have similar electronic and magnetic properties as Ba$_5$As$_3$. The 1D confinement of anionic electrons creates large DOS near $E_F$ and results in the Stoner-type magnetism. The calculated magnetic moments and spin-polarization energies for all 9 $A_5B_3$ compounds are listed in Table 1. The detailed band structures and spin density maps can be found in Supporting Information. All A$_5$B$_3$ electrides have magnetic ground states, and the energy differences between the most stable magnetic state and the NM state are very large which are comparable to those of organic electrides calculated by density functional theory.\cite{PCCP2018} The largest (smallest) spin-polarization energy is found in Ba$_{10}$As$_6$ (Ca$_{10}$Bi$_6$) whose cations and anions have the largest (smallest) electronegativity difference. As is proposed previously, the property of anionic electrons in electrides is related to the electronegativity difference between cations and anions.\cite{electronegativity} Meanwhile, the confined anionic electrons will dominate the magnetism of electrides. That is to say, the electronegativity difference between cations and anions in electrides will affect the magnetic stability. Interestingly, the FM state becomes more energetically favorable than the AFM state with the decreasing electronegativity difference, and the energy of the FM state relative to the AFM$_{\rm zig}$ state changes from 18.4 meV (Ba$_{10}$As$_6$) to $-$74.6 meV (Ca$_{10}$Bi$_6$). For most compounds, the energy difference between ground magnetic states and other magnetic states are several tens of meV per formula unit. These values are in the same order as the well-known magnetic organic electrides, suggesting that the magnetic ground state can be well sustained against perturbation.\cite{PCCP2018,JPCC,PCCP2011} All these results are confirmed by the hybrid functional calculations (see Table S1). The band structures and spin density maps for all these compounds are shown in Figure S2. Furthermore, magnetic stability is checked by applying biaxial strain (data are not shown). It is found that neither the type nor the magnitude of magnetism can be detectably influenced by strain, implying that the magnetism in $A_5B_3$ is robust against different substrates.

As discussed above, the magnetic ground state (FM or AFM) of electride depends on the exchange interaction of the anionic electrons, which is complex and will be studied in the future work. Here it should be noticed that, till now all experimentally characterized magnetic electrides are AFM, therefore, the discovery of electride with FM ground state is of particular significance. Since the magnetic density primarily concentrates on the anionic electrons, ferromagnetic electrides can generate pure spin current with low work function and high conductivity, which both have great potential in spintronics based on low-dimensional magnetism. Because all these materials have been already synthesized in experiments,\cite{experBaAs} electronic and magnetic properties are waiting to be characterized.

\subsection{0D magnetic electrides}

\renewcommand\arraystretch{1.2}\setlength{\tabcolsep}{6pt}
\begin{table}[bp]
	\caption{Calculated magnetic properties for 0D magnetic electrides, where $M_{\rm FM}$ (in unit of $\mu_B$) are the magnetic moments for the FM states and $\Delta E_{\rm FM}$, $\Delta E_{\rm AFM}$ (in units of meV) are the energies of FM, AFM states relative to the NM state. The energy difference between FM and AFM states $\Delta E_{\rm rel}$ are given. For each compound, the ground magnetic state is labeled by *.}
	\label{tb:tb1}
	\begin{tabular}{ccccc}
		\hline
		\hline
		System  &$M_{\rm FM}$  &$\Delta E_{\rm FM}$   &$\Delta E_{\rm AFM}$  &$\Delta E_{\rm rel}$ \\
		\hline
		Sr$_{11}$Mg$_2$Si$_{10}$  &$0.83$   &$-49.7$    &$-49.9$\textsuperscript{*}    &$0.2$ \\
        Ba$_7$Al$_{10}$       &$1.46$   &$-4.5$\textsuperscript{*}     &$28.9$     &$-33.3$ \\
		Ba$_4$Al$_5$          &$0.53$   &$-0.7$\textsuperscript{*}     &$17.6$     &$-18.2$ \\
		\hline
	\end{tabular}\\
\end{table}

\begin{figure} [tbp]
\centering
 \includegraphics[width=0.8\textwidth]{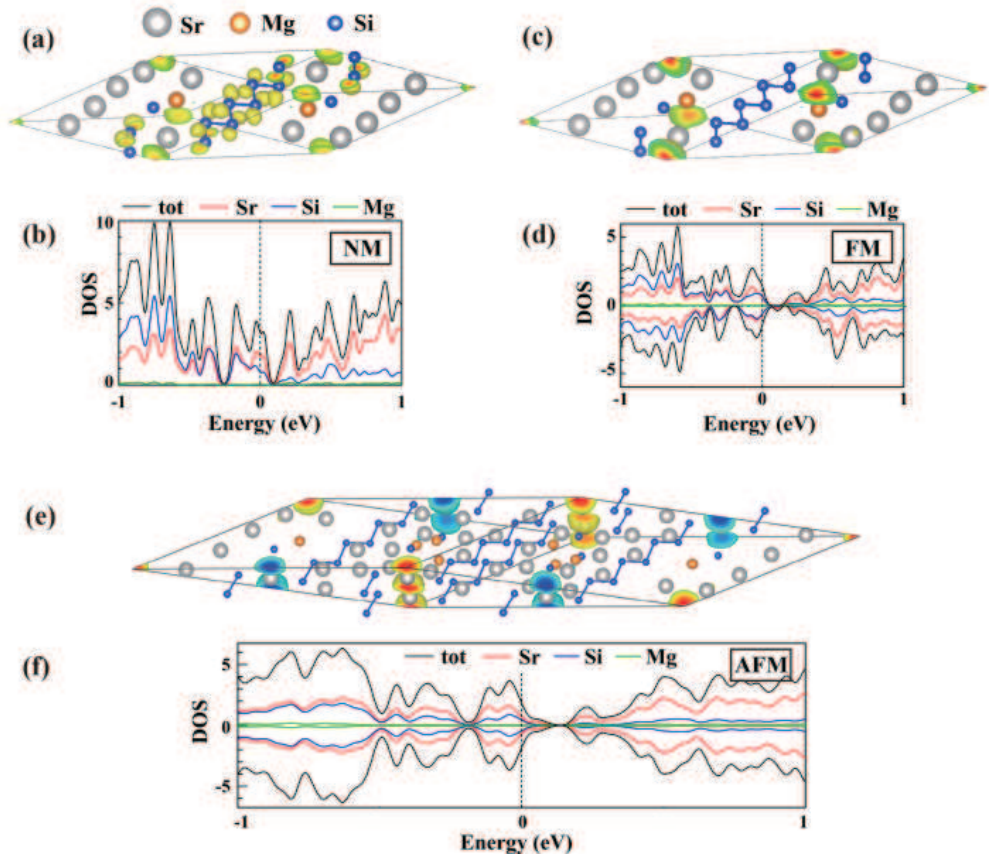}\\
 \caption{(Color online) (a) Partial charge density distribution around $E_F$ ($-$0.2 to 0.2 eV) for Sr$_{11}$Mg$_2$Si$_{10}$. Isosurface is set at 0.02 $e$/\AA$^3$. (b) Total and partial DOS for NM state. (c)(e) Spin density distributions for the FM and AFM states, where the isosurface is at 0.01 $\mu_B$/\AA$^3$ and red (blue) region represents the majority (minority) spin density. (d)(f) Partial DOS for the FM and AFM states. Gray, gold and blue spheres represent Sr, Mg and Si atoms, respectively.}
 \label{fig:Figure4}
\end{figure}

For 0D electrides, three systems, Sr$_{11}$Mg$_2$Si$_{10}$ (space group: $C2/m$), Ba$_7$Al$_{10}$ (space group: $R\bar3m$) and Ba$_4$Al$_5$ (space group: $P6_3/mmc$) are found to be magnetic. The magnetic moments and relative energies are given in Table2, which shows that all 0D electrides have magnetic ground states. The crystal structure of Sr$_{11}$Mg$_2$Si$_{10}$ is complicated, as illustrated in Figure 4a. The partial charge density ($E_F-0.2$ eV < $E$ < $E_F+0.2$ eV) shows that some of the electrons are distributed in the 0D cavities surrounded by Sr atoms, while some of them occupy the atomic orbitals of Si. The partial DOS also reflects that the DOS near $E_F$ is contributed by the orbital electrons of Sr and Si as well as the anionic electrons [Figure 4b]. In addition, a sharp DOS peak is located at $E_F$, indicating the Stoner-type instability. This DOS peak splits and shifts away from $E_F$ in our spin-polarized calculations [Figure 4d,f] for both FM and AFM configurations. The calculated energy difference between FM and NM states is $-$49.7 meV per unit cell, and the magnetic moment is 0.83 $\mu_B$. The energies of FM and AFM states are comparable, and the energy of AFM state is 0.2 meV lower than that of the FM state. Figure 4c and e show the spin density distributions of FM and AFM states. It clearly reveals that the magnetic moment is mainly contributed by the anionic electrons in 0D cavities. Both of the other two 0D magnetic electrides (Ba$_7$Al$_{10}$ and Ba$_4$Al$_5$) show the FM ground state and the spin densities are shown in Figure S3. Note that these three compounds have also been fabricated in experiments.\cite{experSrMgSi,experBaAl}

It is worth noting that the spin-polarization energies for 0D electrides here are relatively small ($-$49.7 meV, $-$4.5 meV and $-$0.7 meV per formula unit for Sr$_{11}$Mg$_2$Si$_{10}$, Ba$_7$Al$_{10}$ and Ba$_4$Al$_5$, respectively), almost an order of magnitude smaller than those for 1D electrides $A_5B_3$. Therefore, these 0D electrides might have weaker magnetic stability against perturbation. It seems to contradict with the conclusion we have drawn in Figure 1 that 1D and 0D electrides are likely to form magnetic ground states. Here we should keep in mind that the conclusion is only suitable for the ideal situation, i.e., the excess electrons (in consideration of the formal valence states of electrides) are completely confined in the cavities and act as anionic electrons. In practice, there are not only anionic electrons but also a certain proportion of orbital electrons around $E_F$ for the electrides. Their hybridizations partially delocalize the anionic electrons, and further affect the $D$($E_F$) to lower the peak. In other words, besides the strong dependence on the confinement dimensionality, the magnetic stability of electrides is also related to the hybridization degree between the anionic electrons and orbital electrons. This effect is particularly pronounced in 0D electrides. Comparing Figure 4 and to Figure 2, it is obvious that the occupation of orbital electrons near $E_F$ has a larger proportion for 0D electride Sr$_{11}$Mg$_2$Si$_{10}$, which weakens its magnetic stability. For 1D electrides, the proportion of orbital electrons around $E_F$ are small and the $D$($E_F$) for nonmagnetic states are large. The inclusion of spin polarization causes the DOS peak to split and shift away from $E_F$, lowering the total energies greatly and stabilizing the magnetic states. In a nut shell, it is the strong hybridization of the anionic electrons and orbital electrons that weakens the magnetic stability in 0D electrides.

\section{Conclusion}
In conclusion, we have proposed that Stoner-type instability can be achieved in low-dimentional electrides due to the quantum confinement effect. Using first-principles calculations, we successfully identified a series of 1D and 0D magnetic electrides, including $A_5B_3$ ($A$ = Ca, Sr, Ba; $B$ = As, Sb, Bi) (1D), Sr$_{11}$Mg$_2$Si$_{10}$ (0D), Ba$_7$Al$_{10}$ (0D) and Ba$_4$Al$_5$ (0D). These electrides have stable magnetic ground states (FM or AFM), with spin-polarization energies up to hundreds of meV. The magnetic moment and spin density are mainly derived from the interstitial anionic electrons. Moreover, we demonstrate that besides the dependence on the confinement dimensionality, the magnetism of electrides is also related to the hybridizations of the anionic electrons and orbital electrons. Such hybridizations delocalize the anionic electrons and weaken the magnetic stability. These ferromagnetic electrides can generate pure spin current with low work function and high conductivity. Our work not only provides a theoretical guidance for the searching of stable magnetic electrides, but also paves a way to the exploration of the magnetic applications in electrides, such as the spin injection in spintronics.

\begin{suppinfo}
Detailed calculation results of band structure for magnetic Ba$_5$As$_3$, magnetic properties calculated by HSE method, band structures and spin density maps for all $A_5B_3$ compounds, and spin density distributions for 0D magnetic electrides.
\end{suppinfo}

\begin{acknowledgement}
The authors thank the support of the Ministry of Science and Technology of China (Grant No. 2016YFA0301001) and the National Natural Science Foundation of China (Grants No. 11674188 and 51788104). This work is supported in part by the Beijing Advanced Innovation Center for Future Chip (ICFC).
\end{acknowledgement}


\end{document}